 \documentclass[preprint ]{aastex} 

\shorttitle{EUVE Spectroscopy of V834 Centauri}
\shortauthors{Mauche}

 \slugcomment{Accepted for publication in 
             {\it Ap.J.\/}                     2002 June 17}

\newcommand{\Mdot}{\dot{M}}
\newcommand{\Mwd  }{M_{\rm wd}}
\newcommand{\Rwd  }{R_{\rm wd}}
\newcommand{\Msun}{{\rm M_{\odot}}}
\newcommand{\lax}{{\lower0.75ex\hbox{ $<$ }\atop\raise0.5ex\hbox{ $\sim$ }}}
\newcommand{\gax}{{\lower0.75ex\hbox{ $>$ }\atop\raise0.5ex\hbox{ $\sim$ }}}

\begin{document}

\title{EUVE Phase-Resolved Spectroscopy of V834 Centauri}

\author{Christopher W.\ Mauche}
\affil{Lawrence Livermore National Laboratory, \\
       L-43, 7000 East Avenue, Livermore, CA 94550; \\
       mauche@cygnus.llnl.gov}

\clearpage 


\begin{abstract}

The {\it Extreme Ultraviolet Explorer\/} ({\it EUVE\/}) satellite was
employed for 5.46 days beginning on 1999 February 9.03 UT to acquire
phase-resolved EUV photometric and spectroscopic observations of the
AM~Her-type cataclysmic variable V834 Centauri. The resulting data
are superior to those obtained by {\it EUVE\/} beginning on 1993 May
28.14 UT because the source was approximately three times brighter, the
observation was four times longer and dithered, and {\it ASCA\/} observed
the source simultaneously. Although we do not understand the EUV light
curves in detail, they are explained qualitatively by a simple model of
accretion from a ballistic stream along the field lines of a tilted
($[\beta , \psi ]\approx [10^\circ , 40^\circ ]$) magnetic dipole
centered on the white dwarf. In 1993 when the EUV flux was lower,
accretion was primarily along the $\varphi\approx\psi\approx 40^\circ $
field line, whereas in 1999 when the EUV flux was higher, accretion took
place over a broad range of azimuths extending from $\varphi\approx
\psi\approx 40^\circ $ to $\varphi\approx 76^\circ $. These changes in
the accretion geometry could be caused by an increase in the
mass-accretion rate and/or the clumpiness of the flow. The 75--140~\AA \
{\it EUVE\/} spectra are well described by either a blackbody or a pure-H
stellar atmosphere absorbed by a neutral hydrogen column density, but
constraints on the size of the EUV emission region and its UV brightness
favor the blackbody interpretation. The mean 1999 EUV spectrum is best
fit by an absorbed blackbody with temperature $kT\approx 17.6$ eV, 
hydrogen column density $N_{\rm H}\approx 7.4 \times 10^{19}~\rm
cm^{-2}$, fractional emitting area $f\approx 10^{-3}$, 70--140~\AA \
$\rm flux\approx 3.0\times 10^{-11}~\rm erg~cm^{-2}~s^{-1}$, and
luminosity $L_{\rm soft}\approx 7.2\times 10^{32}\, (d/100~{\rm
pc})^2~\rm erg~s^{-1}$. The ratio of the EUV to X-ray luminosities is
$L_{\rm soft} /L_{\rm hard}\approx 40$, signaling that some mechanism
other than irradiation (e.g., blob heating)  dominates energy input into
the accretion spot. The 1999 SW hardness  ratio variation can be
explained by minor variations in $kT$ and/or $N_{\rm H}$, but instead of
tracking the SW count rate variation, the hardness ratio variation was
sinusoidal, with a minimum (maximum) when the accretion spot was on the
near (far) side of the white dwarf, consistent with the trend expected
for an atmosphere with an inverted temperature distribution.

\end{abstract}


\clearpage 
 

\section{Introduction}

Polars or AM~Her stars are a class of semidetached binaries composed of
a low-mass main-sequence secondary and a strongly magnetic ($B\approx
10$--100 MG) white dwarf primary. The strong field locks the white dwarf
into corotation and the accreting matter is channeled along the field
lines for much of its trajectory from the secondary's inner Lagrange
point to a spot near the white dwarf magnetic pole. To match boundary
conditions, the flow passes through a strong shock far enough above the
star for the hot ($kT\le kT_{\rm shock}= 3G\Mwd \mu m_{\rm H}/8\Rwd
\sim 20$ keV), post-shock matter to cool and come to rest at the
stellar surface. This plasma cools via bremsstrahlung and line emission
in the X-ray bandpass and cyclotron emission in the optical and
near-IR. Roughly half of this radiation is emitted outward and is observed
directly; the other half is emitted inward, where it is either reflected
or absorbed by the white dwarf surface. In addition to  radiative
heating, the white dwarf surface can be heated by blobs of material which
penetrate to large optical depths before thermalizing  their kinetic
energy. For a mass-accretion rate $\Mdot =10^{16}~\rm g~s^{-1}$ and a
relative spot size $f=10^{-3}$, the white dwarf surface is heated to a
temperature $kT=k(G\Mwd\Mdot /4\pi\sigma f\Rwd ^3)^{1/4}\sim 20$~eV,
hence produces a radiation spectrum which peaks in the EUV.

Recent progress in our understanding of the accretion spots of polars has
come from photometric and spectroscopic observations with {\it EUVE\/}.
The state of our understanding of the {\it EUVE\/} spectra of polars
as of mid-1998 is described by \citet{mau99}, who fit blackbody, pure-H
stellar atmosphere, and solar abundance stellar atmosphere models to
the phase-average {\it EUVE\/} spectra of nine polars with useful data
in the archive. Among the models tested, the blackbody parameterization
gave the best fits to the data, with blackbody temperatures $kT_{\rm
bb}\approx 15$--25~eV and hydrogen column densities $\log N_{\rm H}({\rm
cm^{-2}})\approx 19$--20. Wishing to increase the number of polars with
good signal-to-noise ratio EUV spectra, in 1998 we obtained approval
for a 130 ks {\it EUVE\/} observation of V834~Cen, whose 40 ks archival
spectrum had a {\it peak\/} signal-to-noise ratio of only eight in
0.54~\AA \ bins. Coincidentally, in late 1998 M.~Ishida obtained approval
for an additional {\it ASCA\/} X-ray observation of V834~Cen, so it was
arranged that these observations should be obtained simultaneously. A
description of the X-ray light curves and spectra from the {\it ASCA\/}
observation is provided by \citet{ish99} and \citet{ter01}. Below we
present the analysis of the EUV light curves and spectra from both the
original 1993 and the new 1999 {\it EUVE\/} observations of V834 Cen.

V834 Cen (nee E1405$-$451) is a well-studied polar with an orbital period
$P_{\rm orb}=101.5$ minutes \citep{mas83} and the high ($V\approx14$) and
low luminosity states ($V\approx 17$) typical of this class of binaries.
The magnetic nature of the white dwarf is demonstrated directly by the
linear and circular polarization present in its high states \citep{cro86,
cro89} and the Zeeman absorption features and cyclotron emission features
present in its low states \citep{sch90, fer92}, from which a magnetic
field strength of 23~MG is inferred. The distance $d$, white dwarf mass
$\Mwd $, binary inclination $i$, and the colatitude $\beta $ and azimuth
$\psi $ of the accretion spot are all uncertain at some level, but
\citet{puc90} find $d>77$ pc based on a detection of the secondary in a
near-IR spectrum; \citet{cro99} find $\Mwd=0.54$--$0.64~\Msun $ based
on fits of a {\it Ginga\/} X-ray spectrum, \citet{ram00} finds $\Mwd =
0.64$--$0.68~\Msun $ based on fits of an {\it RXTE\/} X-ray spectrum,
and \citet{sch93} use phase-resolved optical spectra to determine
$\Mwd =0.66^{+0.19}_{-0.16}~\Msun $ for $i={50^\circ }
\, ^{+10^\circ }_{-5^\circ }$; \citet{cro88} advocates $i=45^\circ
\pm 9^\circ $, $\beta =25^\circ\pm 5^\circ $, and $\psi = 40^\circ
\pm 5^\circ $ based on values compiled from the literature.

The plan of this paper is as follows. In \S 2 we describe the {\it 
EUVE\/} observations, in \S 3 we present the {\it EUVE\/} deep survey
(DS) photometer and {\it ASCA\/} count rate light curves, in \S 4 we
present the {\it EUVE\/} short wavelength (SW) spectrometer count rate
and hardness ratio light curves, in \S 5 we present the mean and
phase-resolved SW spectra, in \S 6 we provide a discussion and
interpretation of these data, and in \S 7 we close with a summary of
our results.

\section{Observations}

For a description of the {\it EUVE\/} satellite and instrumentation, refer
to \citet{boy91}, \citet{abb96}, and \citet{sir97}. It is sufficient to
note here that the bandpasses of the {\it EUVE\/} DS photometer and SW
spectrometer are defined by a Lexan/Boron filter and extend from $\approx
70$ to $\approx 180$~\AA , although, as we will see below, interstellar
absorption extinguishes the EUV flux of V834 Cen longward of $\approx
140$~\AA . The first {\it EUVE\/} observation of V834 Cen began on 1993
May 28.14 UT, ran for 1.40 days, and resulted in a total of 42 ks of
exposure, while the second observation began on 1999 February 9.03 UT,
ran for 5.46 days, and resulted in a total of 161 ks of exposure. The
{\it ASCA\/} observation began on 1999 February 9.96 UT, ran for 1.75
days, and resulted in a total of 58 ks of exposure. Subsequent to the
first observation, a procedure was established to increase the
signal-to-noise ratio of {\it EUVE\/} spectra by dithering (delightfully,
to ``dither'' is to ``shiver'' or ``tremble'') the pointing position of
the spacecraft, thereby moving the position of the spectrum on the face
of the detectors and so average over quantum efficiency variations. The
first part of the second observation was erroneously performed undithered,
but after 0.84 days and 25 ks of exposure the spacecraft was reconfigured
to spiral dither. Unfortunately, there is no known record of the optical
brightness of V834 Cen at the time of the 1993 observation, but during
the 1999 observation the source was in a high optical state with
$V\approx 14$ (F.~M.\ Bateson 1999, personal communication)


As is the case for all observations of short-period CVs by low-Earth-orbit
satellites, {\it EUVE\/} observations of V834 Cen are complicated by the
similarity of the satellite's orbital period ($P_{\rm sat}\approx 94.6$
minutes) to the binary's orbital period ($P_{\rm orb}=101.5$ minutes).
This situation is exasperated by the fact that {\it EUVE\/} takes data
only for the $\approx 30$ minutes it is in Earth's shadow. The result is
that the binary phase advances by $\approx 30$\% during each observing
interval, but recedes by $\approx 7$\% for each satellite orbit (i.e.,
$\phi= 0.70$--1.0, 0.63--0.93, 0.56--0.86, etc.). The net result is that
all binary phases are sampled only after 11 satellite orbits or 0.72 days,
while a full cycle is completed only after $(P_{\rm sat}^{-1} -P_{\rm
orb}^{-1})^{-1}\approx 0.96$ days. The 1993 {\it EUVE\/} observation
spans less than two such intervals, while the 1999 observation spans
nearly six.

To phase these data, we assumed the spectroscopic ephemeris of
\citet{sch93}: $T_0({\rm HJD})=2445048.9500(5) + N\times 0.070497518(26)$,
where $N$ is the cycle number, $T_0$ is the blue-to-red zero crossing
of the narrow component of the \ion{He}{2} $\lambda 4686$ and H$\beta $
emission lines in optical spectra of V834 Cen, and the numbers in
parenthesis indicate the uncertainties in the last digits; these errors
result in a phase uncertainty of 0.02 (0.03) cycles at the midpoint of
the 1993 (1999) {\it EUVE\/} observation. For ease of comparison, we
note that Sambruna et al.\ (1991, 1994) used the photometric ephemeris
of \citet{cro86} to phase the {\it EXOSAT\/} data, and at the midpoints
of the 1984, 1985, and 1986 {\it EXOSAT\/} LE + 3000 Lexan observations
and the 1993 and 1999 {\it EUVE\/} observations these ephemerides differ
by $0.15\pm 0.01$, $0.18\pm 0.02$, $0.19\pm 0.02$, $0.34\pm 0.05$, and
$0.46\pm 0.08$ cycles, respectively. The ability of the Schwope et al.\
ephemeris to phase the {\it EXOSAT\/} and {\it EUVE\/} light curves
confirms that that ephemeris is to be preferred.

\section{DS Light Curves}

Assuming the above ephemeris, the 1993 and 1999 background-subtracted DS
count rate light curves are as shown in the upper two panels of Figure~1.
There are clear differences between the 1993 and 1999 DS light curves of
V834 Cen, and although the {\it EUVE\/} light curves are similar to the
1986 and 1985 {\it EXOSAT\/} LE + 3000 Lex light curves, respectively
(cf.\ Fig.~2 of \citealt{sam94}), they also differ in significant ways.
Both the 1993 and 1999 {\it EUVE\/} light curves have a broad maximum
peaking at binary phase $\phi\approx 0.55$, a narrow maximum peaking just
before $\phi\approx 1.0$, a secondary minimum situated between the two
maxima, and a strong asymmetric eclipse centered at $\phi\approx 0.88$.
Compared to the 1999 light curve, the 1993 light curve has a stronger
narrow maximum and a narrower, more symmetric eclipse. Conversely, the
1999 light curve has a more pronounced secondary minimum and a broader,
more asymmetric eclipse.

For completeness, we show in the lower panel of Figure~1 the {\it ASCA\/}
X-ray light curve obtained simultaneously with the 1999 {\it EUVE\/}
observation. This light curve, kindly provided to us by M.~Ishida, is the
summed count rate in the 1.5--10 keV bandpass of the two {\it ASCA\/}
SIS and GIS detectors. The {\it ASCA\/} light curve is seen to differ
significantly from the {\it EUVE\/} light curves: it appears to consist
of a double hump superposed on a constant background. The energy-resolved
{\it ASCA\/} light curves of \citet{ish99} and \citet{ter01} show that
the double hump is actually a broader feature cut at low energies by an
asymmetric absorption feature, which, like the {\it EUVE\/} light curves,
reaches minimum at $\phi\approx 0.88$.

Because V834 Cen was a factor of approximately three times brighter in
the EUV during the 1999 observation, and because its brightness was
observed to dim systematically throughout the observation, we subdivided
the 1999 {\it EUVE\/} observation into six equal-length intervals
spanning the length of the observation, and constructed the six
independent DS light curves shown in Figure~2. Modulo the overall
decrease in the brightness of the source, the light curves from the
various intervals generally follow the mean light curve. It is apparent
from Figure~2 that during the 1999 observation the eclipse was composed
of a stable, narrow (full width $\Delta\phi\approx 0.05$), full eclipse
centered at $\phi\approx 0.88$, and a variable, broad, partial eclipse
centered at $\phi\approx 0.79$. The largest differences between the
various intervals occur at (1) $\phi\approx 0.84$, at the interface
between the partial and full eclipses, (2) $\phi\approx 0.07$, on the
decline from the narrow maximum (compare the first and fifth panels of
Fig.~2), and (3) $\phi\approx 0.17$, in the detailed shape of the
secondary minimum: it appears at some intervals that there is a separate
narrow (full width $\Delta\phi\approx 0.1$) partial eclipse situated in
the valley between the two maxima.

\section{SW Light Curve and Hardness Ratio}

The SW spectrometer supplies an independent measurement of the EUV light
curve of V834 Cen, albeit with lower signal-to-noise ratio because of the
instrument's smaller effective area and effectively higher background.
Mean background-subtracted SW count rate light curves were constructed
in the same manner as those of the DS with the exception that flux was
collected only over the band 75--140~\AA ; shortward of $\approx 75$~\AA
\ the background increases strongly and longward of $\approx 140$~\AA \
there is very little source flux. The resulting 1993 (1999) mean SW 
count rate light curve is consistent [$\rm \chi^2/degree~of~freedom~(dof)
=12.4/24=0.52$  ($\rm \chi^2/dof =110.4/99=1.12$)] with the corresponding
mean DS count rate light curve after scaling the latter by a factor of
0.083 (0.079). The upper panel of Figure~3 shows the mean SW and DS count
rate light curves from the 1999 observation.

To search for spectral variations associated with the variations in the
count rate light curves, we constructed hardness ratio light curves
by calculating the ratio of the background-subtracted SW counts from
75--95~\AA \ to that from 95--140~\AA . The break-point between the
two bandpasses is arbitrary, but 95~\AA \ was chosen because it gives
roughly equal number of counts in the two bandpasses. As a compromise
between phase resolution and signal-to-noise ratio, 40 phase bins were
used to bin the 1999 observation, while the dimmer 1993 observation
could stand no more than 10. The 1993 hardness ratio light curve can be
fit by a constant $0.81\pm 0.06$ with $\rm \chi^2/dof =6.78/9=0.75$,
while the 1999 hardness ratio light curve, shown in the lower panel of
Figure~3, is fit reasonably well by a sinusoidal function $A+ B\, \sin
2\pi (\phi- \phi_0 )$ with $A=1.07\pm 0.02$, $B=0.22\pm 0.02$, 
$\phi_0 =0.19\pm 0.02$, and $\rm \chi^2/dof=46.0/37=1.24$. The fit is
significantly improved ($\Delta\chi ^2=7.6$) even without changing the
fit parameters if the isolated aberrant datum at $\phi=0.36$ is removed
from the fit. The other significant deviations from the fit coincide with
the partial eclipse at $\phi\approx 0.79$ and the secondary minimum at
$\phi\approx 0.17$. Surprisingly, these deviations are {\it downward\/},
in the direction of softer spectra, contrary to the behavior expected
from photoelectric absorption.

\section{SW Spectra}

Ignoring for the moment the hardness ratio variation observed during
the 1999 observation, we constructed mean SW spectra of V834 Cen from
the data obtained during the two epochs of observations. The analysis
procedure was similar to that described in \citet{mau99}, but the spectra
were constructed from the event data and, as above, care was taken to
avoid intervals when (1) the background was high, (2) the source was
occulted by Earth, and (3) the SW detector was off. The spectra were
binned up by a factor of 8 (from $\Delta\lambda = 0.0674$~\AA \ to
0.539~\AA ) to increase the signal-to-noise ratio, to match the spectral
resolution of the SW instrument ($\rm FWHM = 0.5$~\AA ), and to remove
the nonstatistical correlation between neighboring wavelength bins. To
realize the full spectroscopic potential of the 1999 observation, data
from only the dithered portion of that observation (136 ks of exposure
starting on 1999 February 9.88 UT) was used. The resulting mean
background-subtracted SW spectra are shown in the upper panels of
Figure~4, where the significantly higher quality of the 1999 spectrum
is due to the three times longer exposure and three times higher mean
count rate. The gross shape of these count spectra are similar, but
apparent differences are masked by the effective area of the SW
spectrometer, which peaks at 100~\AA \ and falls off at both ends of the
bandpass.

To determine if the differences between these spectra are significant, we
fit the data with a simple model consisting of a blackbody extinguished
at long wavelengths by photoelectric absorption. For the latter, we
used the EUV absorption cross sections of \citet{rum94} for \ion{H}{1},
\ion{He}{1}, and \ion{He}{2} with abundance ratios of 1:0.1:0.01, as is
typical of the diffuse interstellar medium. Table~1 lists the best-fit
values and 90\% confidence intervals of the parameters of these fits,
the absorbed 70--140~\AA \ flux, and the fractional emitting area $f$
and luminosity $L$ assuming a fiducial white dwarf mass $\Mwd=0.7~\Msun $
(hence $\Rwd=7.8\times 10^8$ cm) and a fiducial distance $d=100$ pc
(hence $[\Rwd/d]^2=6.4\times 10^{-24}$). The fractional emitting area
is defined as the area of a circular spot on the white dwarf with open
angle $\theta _{\rm s}$, divided by the full area of the white dwarf:
$f\equiv 2\pi\Rwd ^2 (1-\cos\theta _{\rm s})/ 4\pi\Rwd^2= (1-\cos\theta
_{\rm s}) /2$; hence, the luminosity $L= 4\pi\Rwd ^2f\sigma T^4$ and the
solid angle $\Omega = \pi (\Rwd/d)^2\sin ^2\theta _{\rm s}= 4\pi f(1-f)
(\Rwd/d)^2$. The spectra of the best-fit models are shown superposed on
the data in the upper panels of Figure~4, the residuals of the fits are
shown in the lower panels of Figure~4, and the 68\%, 90\%, and 99\%
confidence contours of the fits are shown in the left panel of Figure~5.
Within the context of the absorbed blackbody model, the mean spectra from
these two epochs differ at the 90\% confidence level, but Figure~5 shows
that this is due to only a minor variation in
$kT$ and/or $N_{\rm H}$.

To test the sensitivity of the fitted and inferred parameters to the
assumed spectral model, we also fit the 1993 and 1999 mean SW spectra of
V834 Cen with that of an absorbed, pure-H, line-blanketed, NLTE, $\log 
g=8$ stellar atmosphere. A grid of 25 such models was calculated with
TLUSTY v195 \citep[][and references therein]{hub95} with temperatures
distributed evenly in the log between 20 and 260 kK (1.7--22.4 eV); 
for the $\chi ^2$ calculations, off-grid spectra were determined by
interpolation. With this model, we obtained the best-fit parameters and
90\% confidence intervals listed in Table~1, and the 68\%, 90\%, and
99\% confidence contours shown in the right panel Figure~5. As measured
by $\rm \chi ^2/dof$, these fits are just as good as the blackbody model
fits, and while the range of absorbing column densities is very similar,
the temperatures are much lower and the fractional emitting areas and
consequently the luminosities are significantly higher. This result is
a consequence of the fact that the EUV bump in the stellar atmosphere
models contains a relatively small fraction of the total luminosity, so
very large spot sizes and hence very large net luminosities are required
to reproduce the observed EUV fluxes. Indeed, the spot sizes are so large
that they violate the geometric limit $f\le 0.5$ over the region of
parameter space shaded dark grey in Figure~5; for the pure-H stellar
atmosphere model to apply, the square of the angular radius must be
significantly larger than the fiducial value of $(\Rwd/d)^2 = 6.4\times
10^{-24}$ (i.e., V834 Cen must be significantly closer than 100 pc).

We next investigated the cause of the hardness ratio variation observed
during the 1999 observation. As a compromise between phase resolution
and signal-to-noise ratio, we accumulated spectra in two phase intervals
centered on the minimum and maximum of the hardness ratio light curve:
$\phi =0.78$--1.08 and $\phi=0.28$--0.58, which we imaginatively call
``soft'' and ``hard,'' respectively. The resulting background-subtracted
SW spectra are shown in the upper panels of Figure~6. We fit these data
with the absorbed blackbody and stellar atmosphere models discussed
above and obtained the best blackbody model fits and residuals shown in
Figure~6, the best-fit parameters and 90\% confidence intervals listed
in Table~1, and the 68\%, 90\%, and 99\% confidence contours shown in
Figure~7. As with the mean 1993 and 1999 spectra, there is significant
overlap in the fit parameters of the 1999 hard and soft spectra, but the
strong correlation between the model parameters results in fits which are
different at significantly greater than the 99\% confidence level. The
largest difference between the hard and soft spectra is in the absorbing
column density; indeed, for both models, the best-fit temperature of the
hard spectrum is {\it lower\/} than that of the soft spectrum: within
the context of these models, the hard spectrum is hard because it is
more strongly absorbed. The differences between the blackbody and stellar
atmosphere model fits to these data are similar to those of the 1993
and 1999 mean spectra: compared to the blackbody models, the stellar
atmosphere models are far cooler, larger, and more luminous. Again,
the geometric constraint $f\le 0.5$ excludes the region of parameter 
space shaded dark grey in Figure~7.

Another constraint imposed on the models is the predicted value of the
UV flux density. While simultaneous UV observations do not exist for
either of our {\it EUVE\/} observations, V834 Cen has been observed
numerous times by the {\it International Ultraviolet Explorer\/} ({\it
IUE\/}) \citep[e.g.,][]{nou83, mar84, tak88, sam91, sam94}. We extracted
all of these spectra from the {\it IUE\/} Newly Extracted Spectra (INES)
data archive (http://ines.laeff.esa.es/) and determined that the maximum
mean flux density in the nominally line-free bandpass at $1300\pm 25$ \AA
\ is $4\times 10^{-14}~\rm erg~cm^{-2}~s^{-1}~\AA ^{-1}$. The intrinsic
UV flux density will be larger than this value due to reddening, but
unfortunately the amount of reddening is uncertain. Given the upper limit
of $N_{\rm H} =2\times 10^{20}~\rm cm^{-2}$ for the absorbing column
density from the various fits to the {\it EUVE\/} spectra, a reasonable
upper limit to the reddening should be $A_V=0.1$, but from the {\it
IUE\/} spectra themselves \citet{nou83} inferred $A_V= 0.40\pm 0.16$;
\citet{mar84} inferred $A_V<0.47$ at the 90\% confidence level;
\citet{sam91} inferred $A_V=0.10^{+0.15}_{-0.07}$; whereas \citet{sam94}
inferred $A_V=0.75$. Given this uncertainty, we consider the constraints
imposed by the model UV flux densities for $A_V\le 1$, 0.3, and 0.1
assuming the extinction curve of \citet{fit99} with $A_V/E_{B-V}=3.1$, as
is typical of the diffuse interstellar medium. The regions of parameter
space excluded by these constraints are shaded medium, light, and lighter
grey, respectively, in Figures~5 and 7. Note that the UV constrains
severely constrain the range of parameter space allowed to the pure-H
models.

\section{Discussion and Interpretation}

In the previous sections we have described the phenomenology of the EUV
light curves and spectra of V834 Cen measured by {\it EUVE\/} in 1993
May and 1999 February. To help understand these data, we note the
following, based on our general understanding of polars and the results
from optical studies of V834 Cen by \citet{ros87}, \citet{cro88, cro89},
and \citet{sch93}. (1) The binary inclination $i\approx 45^\circ\pm
9^\circ $ and the accretion spot colatitude and azimuth are respectively
$\beta\approx 25^\circ\pm 5^\circ $ and $\psi\approx 40^\circ\pm 5^\circ
$. (2) In V834 Cen only one accretion spot is visible for all orbital
phases because the sum of the spot colatitude and binary inclination is
less than $90^\circ $. (3) Because the binary inclination is greater
than the spot colatitude, the accretion stream passes through the line
of sight to the accretion region once per binary revolution. (4) The
narrow and broad components of the optical emission lines of polars are
understood to be due to, respectively, the heated face of the secondary
and the base of the accretion stream. (5) Blue-to-red zero crossing of
the narrow components of optical emission lines occurs at binary
phase $\phi\approx 0$, placing the secondary (white dwarf) on the near
(far) side of the binary at that phase. (6) Maximum blueshift of the
broad components of optical emission lines occurs at $\phi\approx
0.42$, so at that phase the accretion stream points most directly toward
us. (7) The linear polarization spike also occurs at $\phi\approx 0.42$,
so at that phase the accretion column is on the plane of the sky. (8)
We expect that the accretion column will point most directly toward us
approximately $180^\circ $ later, at $\phi\sim 0.92$. The circular
polarization, EUV, and X-ray light curves are all eclipsed near this
phase because of the passage through the line of sight of the (pre-shock)
accretion stream and (post-shock) accretion column.

To illustrate these results, we constructed the graphic shown in the
left panel of Figure~8, which includes the nominal position on the white
dwarf of the accretion spot ({\it filled trapezoid\/}); the 1993 and 1999
DS light curves ({\it grey and black polar histograms, respectively\/});
and the phases of the maximum blueshift of the broad component of optical
emission lines, the spike in linear polarization light curves, and the
dip in circular polarization light curves. It is seen that the primary
eclipse of the EUV light curves occurs when the accretion spot points
toward the observer, while the dip in circular polarization light curves
appears to occur {\it between\/} the primary and secondary eclipses of
the EUV light curves.

To help envision the accretion geometry of V834 Cen, we constructed a
simple model, shown in the right panels of Figure~8, of the path of
material from the secondary's inner Lagrange point to the white dwarf
surface in the vicinity of the upper magnetic pole. The model consists
of the ballistic stream for a nonmagnetic semidetached binary and the
field lines of a tilted magnetic dipole centered on the white dwarf for
an orbital period $P_{\rm orb}= 101.5$ minutes, mass ratio $q=5.08$
(determining the trajectory of the ballistic stream), white dwarf mass
$\Mwd = 0.66~\Msun $ (determining the white dwarf radius $\Rwd =8.1
\times 10^8$ cm), magnetic colatitude $\beta = 10^\circ $, and magnetic
azimuth $\psi = 40^\circ $. Field lines are drawn for azimuthal angles 
$\varphi = 0^\circ , 10^\circ , 20^\circ , \ldots , \psi +90^\circ $.
If material in the ballistic stream makes it beyond $\varphi\approx\psi
+90^\circ\approx 130^\circ $, it will accrete preferentially onto the
lower magnetic pole, hence disappears from consideration. In this simple
model, the material lost by the secondary travels along the ballistic
stream until it is threaded by the white dwarf magnetic field; it then
leaves the orbital plane and follows the magnetic field lines down to the
white dwarf surface in the vicinity of the magnetic poles. In reality, the
trajectory of the accreting material will be affected by the magnetic
field and the magnetic field will be distorted by the accreting material;
our simple model is a first-order approximation of the path of the
accreting material which ignores these complications.

In the right panels of Figure 8 we see that if material accretes onto
the white dwarf from the full range of possible azimuthal angles, it
will produce a long, thin, arc-shaped accretion region on the white dwarf
surface at the footpoints of the magnetic field lines. Details of the
accretion region are more apparent in Figure~9, which shows the white
dwarf and magnetic field lines for a binary inclination $i=50^\circ $
and binary phases $\phi = 220^\circ , 230^\circ , 240^\circ , \ldots ,
360^\circ $ ($\phi = 0.61$--1.0). For the assumed parameters, the
accretion region extends about $90^\circ $ in azimuth and has a variable
offset of about $20^\circ $ from the magnetic pole, determined by the
varying distance (a maximum of $37.5~\Rwd $ for $\varphi =0^\circ $ and a
minimum of $5.9~\Rwd $ for $\varphi=130^\circ $) to the ballistic stream
from the white dwarf. Accretion along the $\varphi =\psi =40^\circ $
field line (the thick  black curves in Figs.~8 and 9) will produce an
accretion spot at $[\beta , \psi ] =[26^\circ , 40^\circ ]$, consistent
with the values advocated by \citet{cro88}; the other model parameters
are consistent with the radial velocity solution of \citet{sch93}.

\subsection{EUV Light Curves}

Whether a spot or an arc, the flux at Earth from the accretion region
will vary with binary phase because of the varying projected surface area 
of the accretion region and the phase-dependent obscuration by infalling
material. Ignore for the moment obscuration and consider the simplest
possible model of the EUV light curves of V834 Cen, that of a small spot
``painted'' on the white dwarf surface: its projected surface area hence
light curve varies as $\sin i\,\sin\beta\, \cos\, (\phi +\psi) + \cos
i\,\cos \beta $. Such a simple model fails to reproduce even the upper
envelope of the EUV light curves of V834 Cen, and although linear a
superposition of such functions (i.e., multiple spots) can reproduce
some aspects of the data, the fits are not unique. Consider next an
empirical model, using the EUV light curve of AM~Her \citep{pae96} as
a template to explain the 1999 EUV light curve of V834 Cen. Aligning
the two light curves on the respective linear polarization ephemerides,
the AM~Her light curve reproduces well the V834 Cen light curve between
binary phases $\phi\approx 0.0$--0.17. By shifting the AM~Her light curve
by $\Delta\phi = 0.5$, it reproduces well the V834 Cen light curve between
binary phases $\phi\approx 0.17$--0.6. From this exercise, one is led
to imagine that the EUV emission region of V834 Cen is composed of two
spots separated in azimuth by about $180^\circ $: the first spot produces
the falling portion of the V834 Cen light curve between $\phi\approx
0.0$--0.17, while the second spot produces the rising portion of the
light curve between $\phi\approx 0.17$--0.6; the remaining portion of
the light curve is affected by the primary and secondary eclipses. 
Independent support for two accretion spots is provided by \citet{cro89},
who advocates two magnetic poles separated in azimuth by about $230^\circ
$ to explain the occasional presence of two waves in the positional angle
of the linearly polarized optical flux of V834 Cen. Such a two-spot model
allows us to understand two features of the EUV light curves of V834 Cen.
First, the secondary minimum at $\phi\approx 0.17$ discussed in \S 3 can
be understood as being the interface between the falling portion of the
light curve of the leading spot and the rising portion of the trailing
spot. Second, the change between 1993 and 1999 in the relative strengths
of the EUV light curves between phases $\phi \approx 0.0$--0.17 and
$\phi\approx 0.17$--0.6 (cf.\ the top panel of Fig.~1) can be understood
if the leading (trailing) accretion spot was brighter in 1993 (1999).
Unfortunately, this model predicts that the accretion spots lie at
$\phi\approx 0.05$ and $\phi\approx 0.55$, which appears not to be the
case. These considerations teach us the folly of attempting to understand
in detail the EUV light curves of V834 Cen (or polars in general, see
\citealt{sir98}). 

Returning to the discussion of the previous section, we expect that if
the ballistic stream is threaded by the white dwarf magnetic field over
a {\it narrow\/} range of azimuths, the infalling material will form a
narrow stream leading down to a small spot on the white dwarf surface,
whereas if the ballistic stream is threaded by the magnetic field over a
{\it broad\/} range of azimuths, the infalling material will form an
``accretion curtain'' (as it called in intermediate polars) leading down
to an extended arc on the white dwarf surface. Narrow streams are implied
by the optical eclipse studies of HU~Aqr and UZ~For by \citet{har99} and 
\citet{kub00}, respectively, but an extended accretion region is invoked
by \citet{cro89} in V834 Cen to explain changes in the shape of the
optical light curve and the changing morphology of the linear polarization
and position angle curves. We propose that the EUV light curves of V834
Cen are explained by accretion over a narrow range of azimuths in 1993
and a broad range of azimuths in 1999. Specifically, we propose that in
1993 accretion was primarily along the $\varphi\approx\psi\approx
40^\circ $ field line, resulting in a narrow eclipse of a small accretion
spot, while in 1999 accretion was more spread out in azimuth: the primary
accretion channel remained the $\varphi\approx\psi\approx40^\circ $ field
line, but field lines $\approx 36^\circ $ further along the ballistic
stream became active, resulting in the secondary eclipse observed at
$\phi\approx 0.79$. Happily, the proposed shift of the accretion rate
further along the ballistic stream qualitatively explains the reduction
in the relative strength of the EUV light curve between $\phi\approx
0.0$--0.17. What caused the changes in the accretion geometry between
1993 and 1999? One option is a higher accretion rate, since in 1999 the
EUV flux was approximately three times higher than in 1993. Another option
is enhanced clumping of the flow, since greater clumping will result in
magnetic threading over a broader range of azimuths {\it and\/} more
efficient heating of the accretion spot as blobs of material crash into
the stellar surface.

\subsection{EUV Spectra}

Although we have appealed to obscuration by the pre- and post-shock
accretion flow to explain the primary and secondary eclipses in the EUV
light curves of V834 Cen, Figure~3 makes clear that this obscuration
cannot be due to photoelectric absorption. First, the amplitude of the
SW hardness ratio variation is too small. Assuming that the mean EUV
spectrum of V834 Cen is an absorbed blackbody with $kT=17.6$ eV and
$N_{\rm H}=7.4\times 10^{19}~\rm cm^{-2}$ (cf.\ Table~1 and Fig.~7), the
SW hardness ratio equals 1.1, as observed. Doubling $N_{\rm H}$ decreases
the SW count rate by a factor of $\approx 10$, but the SW hardness ratio
increases by a factor of $\approx 4$. In contrast, during the 1999
observation, the SW count rate varied by a factor of $\gax 10$ while the
hardness ratio varied by only $\approx 20$\%. Similar results apply even
if H and He are singly ionized, and only the \ion{He}{2} bound-free
opacity is left to absorb the EUV flux from the accretion spot. It is
actually quite easy to remove the photoelectric opacity of all these
ions: in a collisionally ionized plasma, H is fully ionized for
$T\gax 25$ kK and He is fully ionized for $T\gax 100$ kK. Above that 
temperature, only Thomson opacity is available to reduce the EUV flux,
and an electron column density $N_{\rm e}= -\ln(0.1)/\sigma _{\rm T} =
3.5\times 10^{24}~\rm cm^{-2}$ is required to reduce the EUV flux by a
factor of ten. Second, the SW hardness ratio variation does not track the
SW count rate variation. Instead, at least during the 1999 observation
(see Fig.~3), the hardness ratio variation was sinusoidal, with a minimum
(maximum) when the accretion spot was on the near (far) side of the white
dwarf. Evidently, as the accretion spot rotates to the far side of the
white dwarf, its spectrum systematically hardens. Figure~7 shows that the
detailed spectral fits indicate that this change is best explained by
decreasing $kT$ {\it and\/} increasing $N_{\rm H}$, but within the 68\%
contours this can be accomplished simply by increasing $N_{\rm H}$, and
within the 90\% contours this can be accomplished simply by increasing
$kT$. The last option implies that the accretion region is harder when
observed at shallow viewing angles. This is the trend expected for an
accretion spot subjected to irradiation by the million-degree plasma in
the accretion column \citep{tee94}, although the affect is predicted to
be small in the 75--140~\AA \ bandpass and the large ratio of hard to
soft X-ray luminosities (see below) argues that irradiation does not
dominate the energy deposition into the accretion spot.

The mean 1993 and 1999 and the phase-resolved 1999 {\it EUVE\/} spectra
of V834~Cen can be fit by a blackbody or a pure-H stellar atmosphere model
absorbed by a neutral hydrogen column density with the parameters listed
in Table~1. Compared to the blackbody models, the stellar atmosphere
models are cooler, larger, and more luminous. In fact, as shown by
Figures~5 and 7, the stellar atmosphere models are {\it too\/} big and
{\it too\/} luminous (too bright in the UV) to explain the observations
of V834 Cen. First, the stellar atmosphere model fails to account for
the EUV spectrum acquired during the hard phase of the 1999 observation
because the required angular size is larger than that of the white dwarf.
Second, the stellar atmosphere model fails to account for the mean 1999
EUV spectrum because it produces too much flux in the UV unless the
reddening $A_V\ge 1$. Similar problems are met applying this model to
other polars \citep{mau99}, so we conclude that the pure-H stellar
atmosphere model cannot in general explain the EUV of these magnetic CVs.

Finally, having {\it simultaneously\/} measured the EUV and X-ray spectra
of V834 Cen in 1999, we are in a position to determine the ratio of the
luminosities of the accretion column and accretion spot. Table~1 shows
that the absorbed blackbody fit to the mean 1999 EUV spectrum yields
best-fit parameters $kT\approx 17.6$ eV, $N_{\rm H} \approx 7.4\times
10^{19}~\rm cm^{-2}$, fractional emitting area $f\approx 10^{-3}$,
70--140~\AA \ $\rm flux\approx 3.0\times 10^{-11}~\rm erg~cm^{-2}~s^{-1}$,
and luminosity $L_{\rm soft}\approx 7.2\times 10^{32}\, (d/100~{\rm
pc})^2~\rm erg~s^{-1}$. The averaged 2--10 keV flux measured by {\it
ASCA\/} was $1.5\times 10^{-11}~\rm erg~cm^{-2}~s^{-1}$ \citep{ish99},
implying $L_{\rm hard}\approx 1.8\times 10^{31}\, (d/100~{\rm pc})^2~\rm
erg~s^{-1}$, hence $L_{\rm soft}/L_{\rm hard}\approx 40$ (to within a
factor of $\approx 2$). This imbalance between the soft accretion spot
and hard accretion column luminosities is the famous ``soft X-ray
problem,'' and signals that some mechanism other than irradiation (e.g.,
blob heating) dominates energy deposition into the accretion spot.

\section{Summary}

We have described {\it EUVE\/} observations of V834 Cen obtained in 1993
May and 1999 February. The 1999 data are superior to those obtained in
1993 because the source was approximately three times brighter, the
observation was four times longer and dithered, and {\it ASCA\/} observed
the source simultaneously. Although we do not understand the EUV light
curves in detail, they are explained qualitatively by a simple model
of accretion from a ballistic stream along the field lines of a tilted
($[\beta , \psi ]\approx [10^\circ , 40^\circ ]$) magnetic dipole
centered on the white dwarf. In 1993 when the EUV flux was lower,
accretion was primarily along the $\varphi\approx\psi\approx 40^\circ $
field line, whereas in 1999 when the EUV flux was higher, accretion
took place over a broad range of azimuths extending from $\varphi
\approx\psi\approx 40^\circ $ to $\varphi\approx 76^\circ $. These
changes in the accretion geometry could be caused by an increase in the
mass-accretion rate and/or the clumpiness of the flow. The 75--140~\AA \
{\it EUVE\/} spectra are well described by either a blackbody or a pure-H
stellar atmosphere absorbed by a neutral hydrogen column density, but
constraints on the size of the EUV emission region and its UV brightness
favor the blackbody interpretation. The mean 1999 EUV spectrum is best
fit by an absorbed blackbody with temperature $kT\approx 17.6$ eV,
hydrogen column density $N_{\rm H}\approx 7.4\times 10^{19}~\rm cm^{-2}$,
fractional emitting area $f\approx 10^{-3}$, 70--140~\AA \ $\rm flux
\approx 3.0\times 10^{-11}~\rm erg~cm^{-2}~s^{-1}$, and luminosity
$L_{\rm soft}\approx 7.2\times 10^{32}\, (d/100~{\rm pc})^2~\rm
erg~s^{-1}$. The ratio of the EUV to X-ray luminosities is $L_{\rm soft}
/L_{\rm hard} \approx 40$, signaling that some mechanism other than
irradiation (e.g., blob heating)  dominates energy input into the
accretion spot. The 1999 SW hardness ratio variation can be explained by
minor variations in $kT$ and/or $N_{\rm H}$, but instead of tracking the
SW count rate variation, the hardness ratio variation was sinusoidal,
with a minimum (maximum) when the accretion spot was on the near (far)
side of the white dwarf, consistent with the trend expected for an
atmosphere with an inverted temperature distribution.

\acknowledgments

Special thanks are due to F.\ Bateson, Director of the VSS/RASNZ, for
helping us plan the 1999 {\it EUVE\/} and {\it ASCA\/} observations of
V834 Cen by instituting a program of optical observations of the source,
and then regularly supplying us with visual magnitudes estimates. We
thank M.~Ishida for graciously supplying the {\it ASCA\/} data shown in
Figure 1, S.~Howell for kindly supplying the ballistic stream trajectory
shown in Figure 8, and S.~Howell, K.~Mukai, and P.~Szkody for helpful
discussions. Dispensation to schedule the {\it EUVE\/} observations
during the {\it ASCA\/} observations was kindly granted by {\it EUVE\/}
Project Scientist R.~Malina. The {\it EUVE\/} observations were scheduled
and performed by {\it EUVE\/} Science Planner M.~Eckert, the staff of the
{\it EUVE\/} Science Operations Center at CEA, and the Flight Operations
Team at Goddard Space Flight Center. We acknowledge with thanks the
use in this research of INES data from the {\it IUE\/} satellite. The
manuscript was improved by the comments and suggestions of the anonymous
referee. This work was performed under the auspices of the U.S.\
Department of Energy by University of California Lawrence Livermore
National Laboratory under contract No.\ W-7405-Eng-48.
 
\clearpage 
 

\clearpage 


\begin{deluxetable}{lcccccc}
\footnotesize
\tablecolumns{7} 
\tablewidth{0pc} 
\tablenum{1}
\tablecaption{Blackbody and Pure-H Model Fit Parameters\tablenotemark{a}}
\tablehead{
\colhead{} &
\colhead{$kT$} &
\colhead{$N_{\rm H}$} &
\colhead{70--140 Flux} &
\colhead{} &
\colhead{$L$} &
\colhead{} \\
\colhead{Phase} &
\colhead{(eV)} &
\colhead{$(10^{19}~\rm cm^{-2})$} &
\colhead{$(10^{-11}~\rm erg~cm^{-2}~s^{-1})$} &
\colhead{$\log f$} &
\colhead{$(10^{33}~\rm erg~s^{-1})$} &
\colhead{$\chi^2$/dof}}
\startdata
\cutinhead{Blackbody Model}
\hbox to 1.0in{1993 Mean\leaders\hbox to 0.5em{\hss.\hss}\hfill}& $14.5^{+9.0}_{-4.7}$& $7.7^{+4.4}_{-3.5}$& $0.9^{+3.2}_{-0.5}$& $-2.8^{+2.6}_{-2.0}$& $0.59^{+54.5}_{-0.55}$& 126/119\\
\hbox to 1.0in{1999 Mean\leaders\hbox to 0.5em{\hss.\hss}\hfill}& $17.6^{+1.6}_{-1.6}$& $7.4^{+0.8}_{-0.7}$& $3.1^{+0.8}_{-0.5}$& $-3.0^{+0.4}_{-0.4}$& $0.72^{+0.65}_{-0.28}$& 179/119\\
\hbox to 1.0in{1999 Hard\leaders\hbox to 0.5em{\hss.\hss}\hfill}& $15.9^{+2.4}_{-1.9}$& $9.1^{+1.3}_{-1.1}$& $4.3^{+1.7}_{-1.1}$& $-2.3^{+0.7}_{-0.6}$& $2.6^{+5.0}_{-1.6}$&    129/119\\
\hbox to 1.0in{1999 Soft\leaders\hbox to 0.5em{\hss.\hss}\hfill}& $17.6^{+4.4}_{-3.3}$& $6.6^{+1.7}_{-1.5}$& $2.1^{+1.3}_{-0.7}$& $-3.3^{+1.0}_{-0.9}$& $0.39^{+1.29}_{-0.26}$& 138/119\\
\cutinhead{Pure-H Model}
\hbox to 1.0in{1993 Mean\leaders\hbox to 0.5em{\hss.\hss}\hfill}& $\phn 2.8^{+7.0}_{-0.8}$& $8.2^{+1.0}_{-3.6}$& $0.9^{+69.0}_{-0.9}$&     $+0.7^{+1.0}_{-3.7}$& $2.4^{+3.2}_{-2.3}$& 127/119\\
\hbox to 1.0in{1999 Mean\leaders\hbox to 0.5em{\hss.\hss}\hfill}& $\phn 4.5^{+1.3}_{-1.0}$& $8.1^{+0.7}_{-0.9}$& $3.1^{+22.4}_{-2.7}$&     $+0.1^{+0.4}_{-0.8}$& $3.7^{+0.4}_{-1.9}$& 182/119\\
\hbox to 1.0in{1999 Hard\leaders\hbox to 0.5em{\hss.\hss}\hfill}& $\phn 3.5^{+1.5}_{-1.0}$& $9.8^{+0.9}_{-1.2}$& $4.3^{+118}_{-4.1}$&      $+0.7^{+0.7}_{-0.7}$& $5.7^{+2.8}_{-0.8}$& 129/119\\ 
\hbox to 1.0in{1999 Soft\leaders\hbox to 0.5em{\hss.\hss}\hfill}& $\phn 4.6^{+3.8}_{-1.9}$& $7.3^{+1.5}_{-1.8}$& $2.1^{+337}_{-2.0}$&      $-0.1^{+1.1}_{-2.1}$& $2.7^{+1.3}_{-2.5}$& 139/119\\
\enddata
\tablenotetext{a}{Best-fit values and 90\% confidence intervals.}
\end{deluxetable}

\clearpage 


\begin{figure}
\figurenum{1}
\epsscale{0.5138}
\plotone{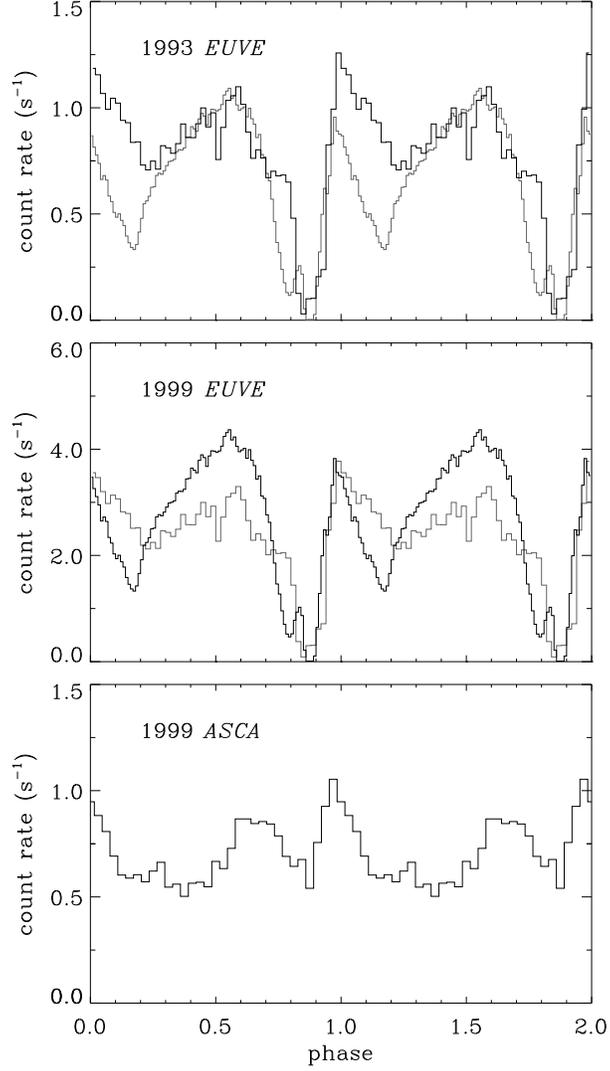}
\caption{{\it EUVE\/} DS and {\it ASCA\/} SIS+GIS 1.5--10 keV count rate
light curves. In the upper (middle) panel, the DS count rate light curve
from the 1999 (1993) observation [scaled by 0.25 (3.0)] is shown by the
grey histogram.
$1\, \sigma $ error vectors of the 1993 and 1999 {\it EUVE\/} and 
{\it ASCA\/} light curves are typically 0.025--0.040, 0.030--0.055, and
0.015--$0.030~\rm counts~s^{-1}$, respectively.}
\end{figure}

\begin{figure}
\figurenum{2}
\epsscale{0.9725}
\plotone{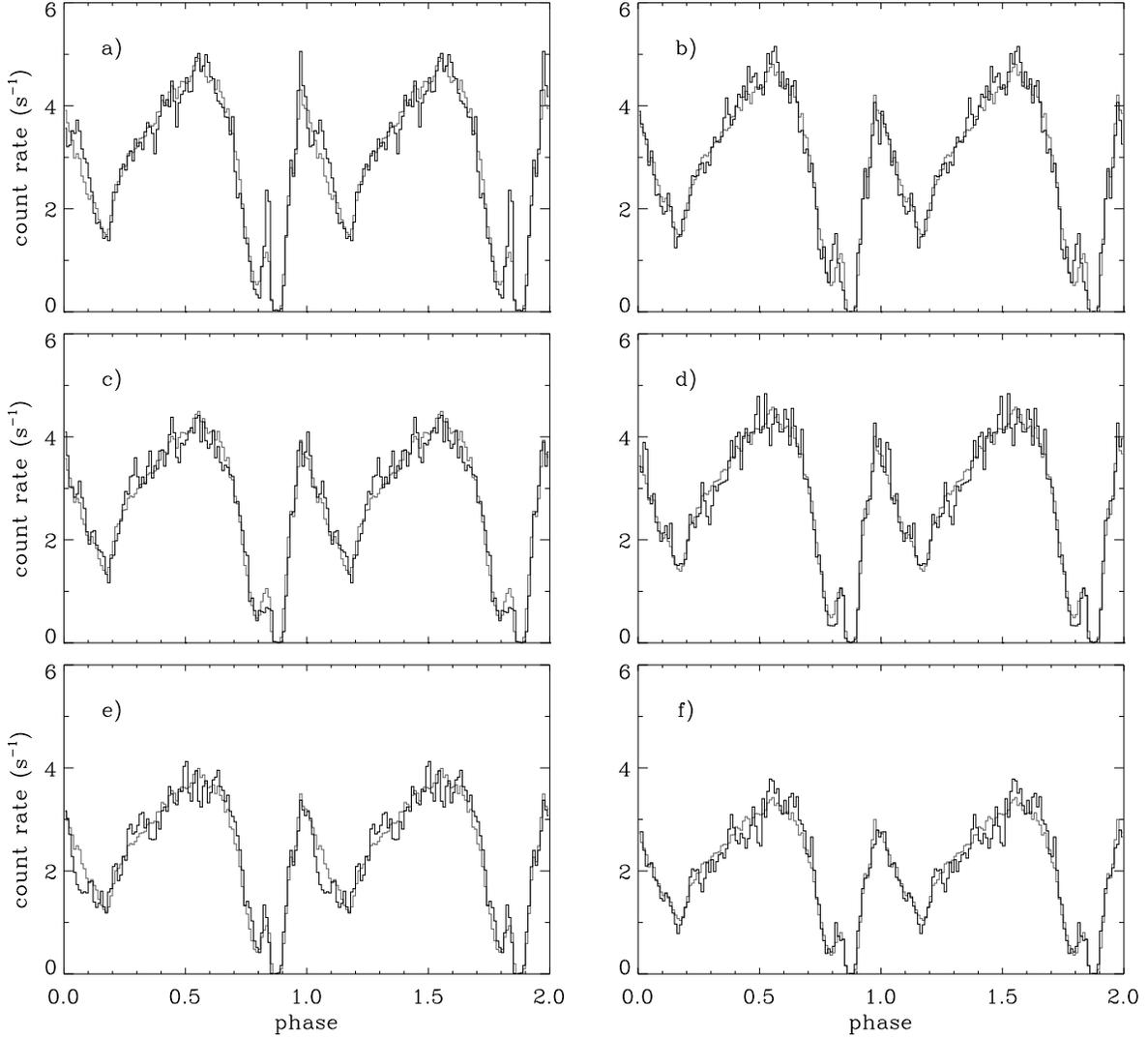}
\caption{DS count rate light curves from the 1999 observation.
Individual light curves are shown by the dark histograms, while the mean
DS light curve (scaled by 1.13, 1.10, 1.03, 1.05, 0.91, 0.78) is shown by
the grey histograms. $1\,\sigma$ error vectors for the individual light
curves are typically 0.05--$0.15~\rm counts~s^{-1}$.}
\end{figure}

\begin{figure}
\figurenum{3}
\epsscale{0.5138}
\plotone{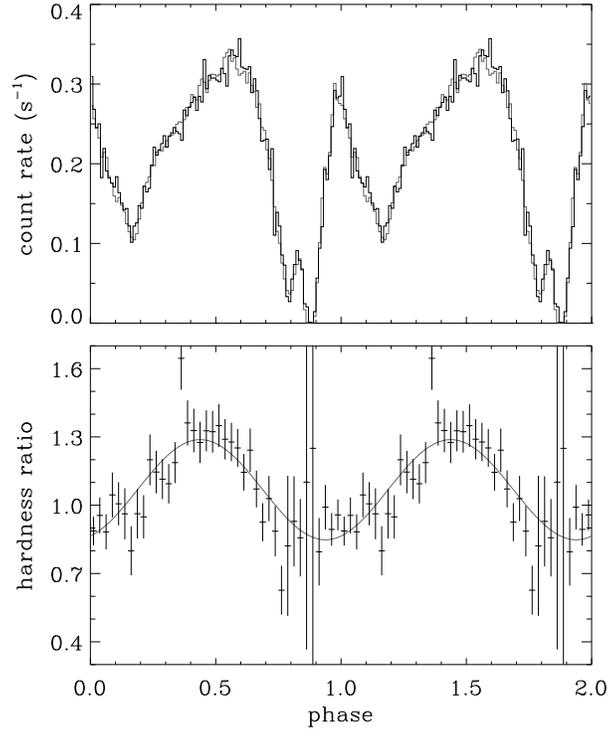}
\caption{{\it Upper panel:\/} SW and DS count rate light curves from
the 1999 observation. SW light curve is shown by the dark histogram,
while the DS light curve (scaled by 0.079) is shown by the grey
histogram. $1\, \sigma$ error vector of the SW light curve is typically
0.01--$0.02~\rm counts~s^{-1}$. {\it Lower panel:\/} SW hardness ratio
light curve $H/S$, where $H$ is the count rate in the 75--95~\AA \ 
waveband and $S$ is the count rate in the 95--140~\AA \ waveband. Smooth
curve is the fit $H/S=1.07 + 0.22\, \sin 2\pi (\phi -0.19)$.}
\end{figure}

\begin{figure}
\figurenum{4}
\epsscale{0.9817}
\plotone{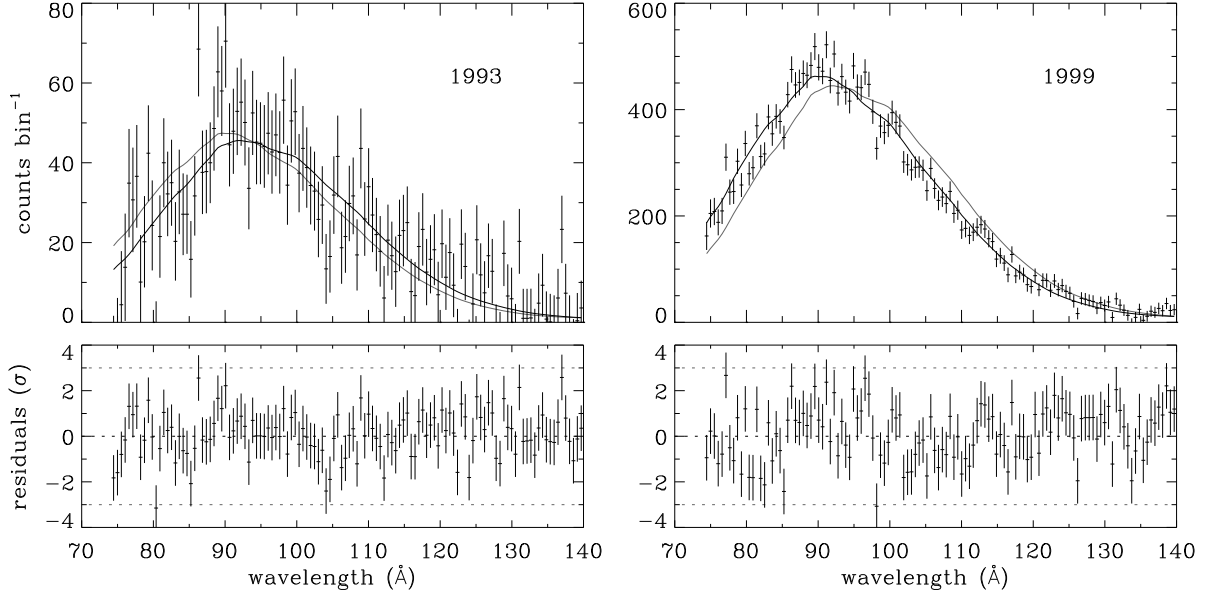}
\caption{{\it Upper panels:\/} Mean SW spectra from the 1993 ({\it
left\/}) and 1999 ({\it right \/}) observations. In the left (right)
panel the  dark curve is the best-fit blackbody spectrum of the 1993
(1999) observation, and the grey curve is the best-fit spectrum of the
1999 (1993) observation scaled by 0.10 (9.8). {\it Lower panels:\/}
Residuals to the best-fit spectra.}
\end{figure}

\begin{figure}
\figurenum{5}
\epsscale{0.9908}
\plotone{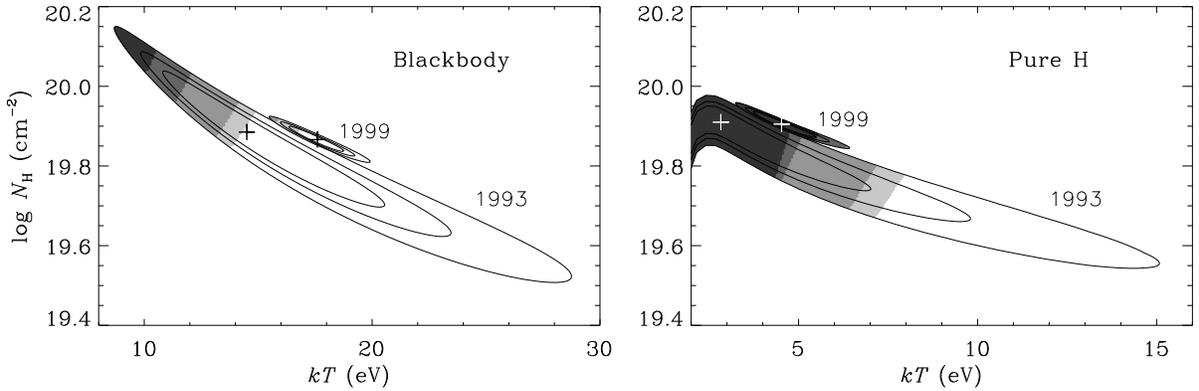}
\caption{Best fit and 68\%, 90\%, and 99\% confidence contours of the
absorbed blackbody ({\it left panel\/}) and pure-H ({\it right panel\/})
model fits of the mean SW spectra from the 1993 and 1999 observations.
90\% confidence intervals of the fit parameters are listed in Table~1.
Regions of parameter space shaded dark, medium, light, and lighter grey
are excluded by the constraints $f\le 0.5$, $A_V\le 1$, $A_V\le 0.3$, and
$A_V\le 0.1$, respectively.}
\end{figure}

\begin{figure}
\figurenum{6}
\epsscale{0.9908}
\plotone{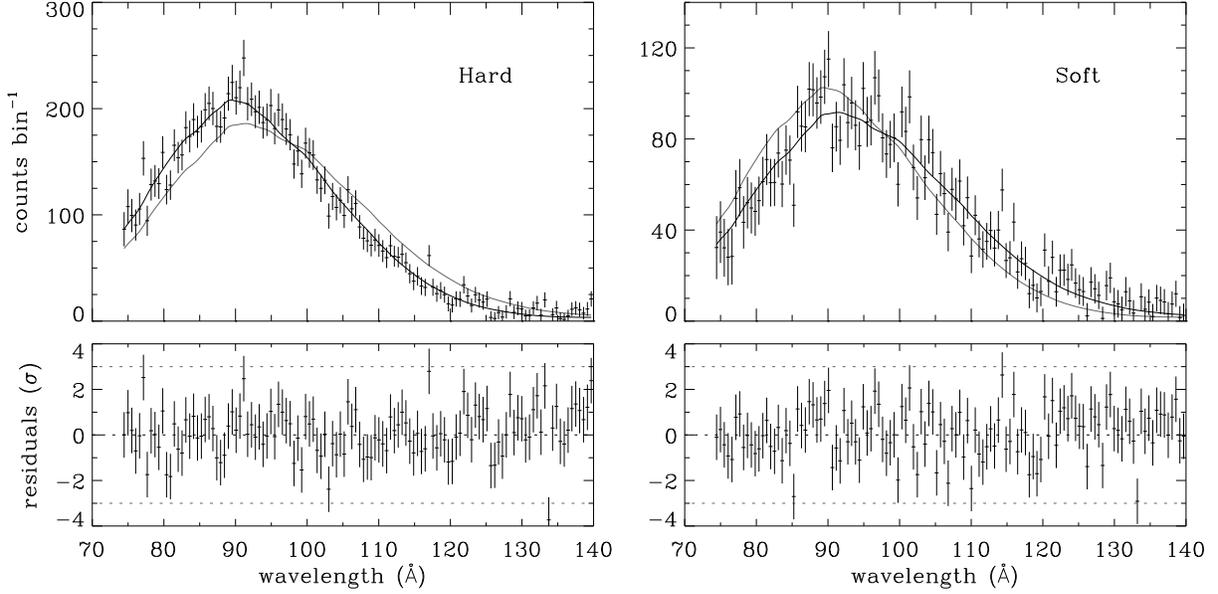}
\caption{{\it Upper panels:\/} SW spectra from the hard ({\it left\/})
and soft ({\it right \/}) phases of the 1999 observation. In the left
(right) panel the dark curve is the best-fit blackbody spectrum of the
hard  (soft) phase, and the grey curve is the best-fit spectrum of the
soft (hard) phase scaled by 2.0 (0.5). {\it Lower panels:\/} Residuals
to the best-fit spectra.}
\end{figure}

\begin{figure}
\figurenum{7}
\epsscale{1.0}
\plotone{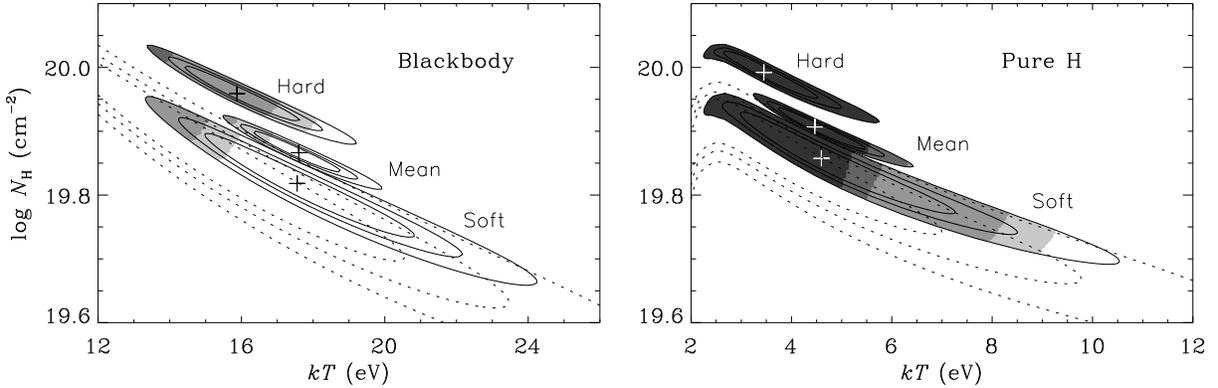}
\caption{Best fit and 68\%, 90\%, and 99\% confidence contours of the
absorbed blackbody ({\it left panel\/}) and pure-H ({\it right panel\/})
model fits of the mean, hard, and soft SW spectra from the 1999
observation. Confidence contours of the mean spectrum from the 1993
observation are shown by the dotted curves. 90\% confidence intervals
of the fit parameters are listed in Table~1. Regions of parameter space
shaded dark, medium, light, and lighter grey are excluded by the
constraints $f\le 0.5$, $A_V\le 1$, $A_V\le 0.3$, and $A_V\le 0.1$,
respectively.}

\end{figure}

\begin{figure}
\figurenum{8}
\plotone{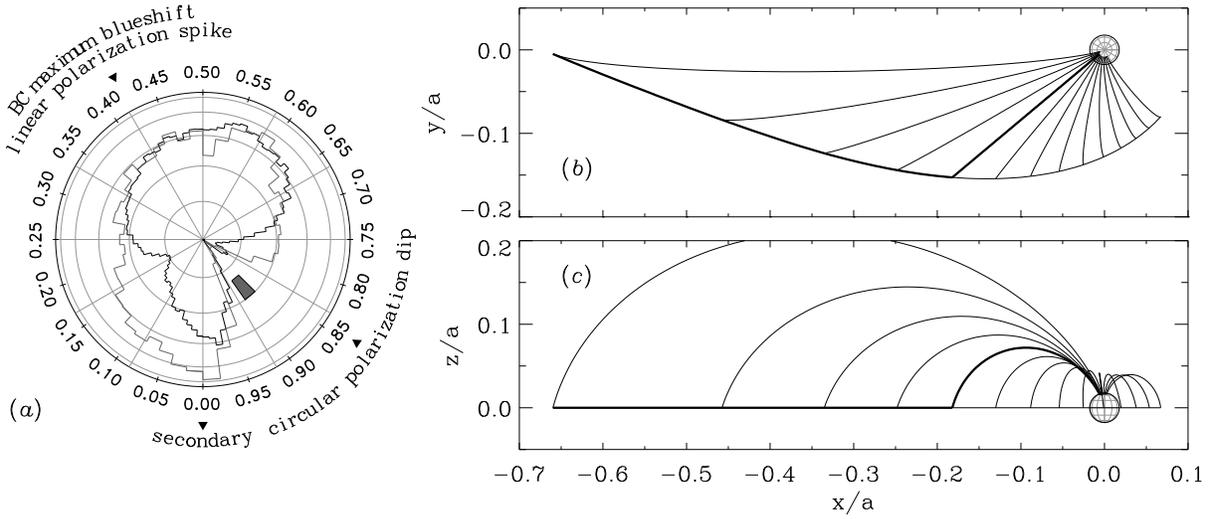}
\caption{({\it a\/}) Schematic diagram of the V834 Cen white dwarf
showing the nominal position of the accretion spot ({\it filled
trapezoid\/}); the 1993 and 1999 DS light curves ({\it grey and black
polar histograms, respectively\/}); and the phases of the maximum
blueshift of the broad component of optical emission lines, the spike
in the linear polarization light curve, and the dip in the circular
polarization light curve. Model of the white dwarf, ballistic stream,
and magnetic field lines as seen from above ({\it b\/}) and from the
side ({\it c\/}) for binary phase $\phi=0.75$. Field lines for a tilted
($[\beta , \psi ] =[10^\circ , 40^\circ ]$) centered magnetic dipole are
drawn for azimuthal angles $\varphi =0^\circ , 10^\circ, 20^\circ , 
\ldots , \psi  +90^\circ $. Dominant accretion path is indicated by the
bold curve. Coordinates are measured relative to the semi-major axis
$a=4.6\times 10^{10}$ cm.}
\end{figure}

\begin{figure}
\figurenum{9}
\epsscale{0.75}
\plotone{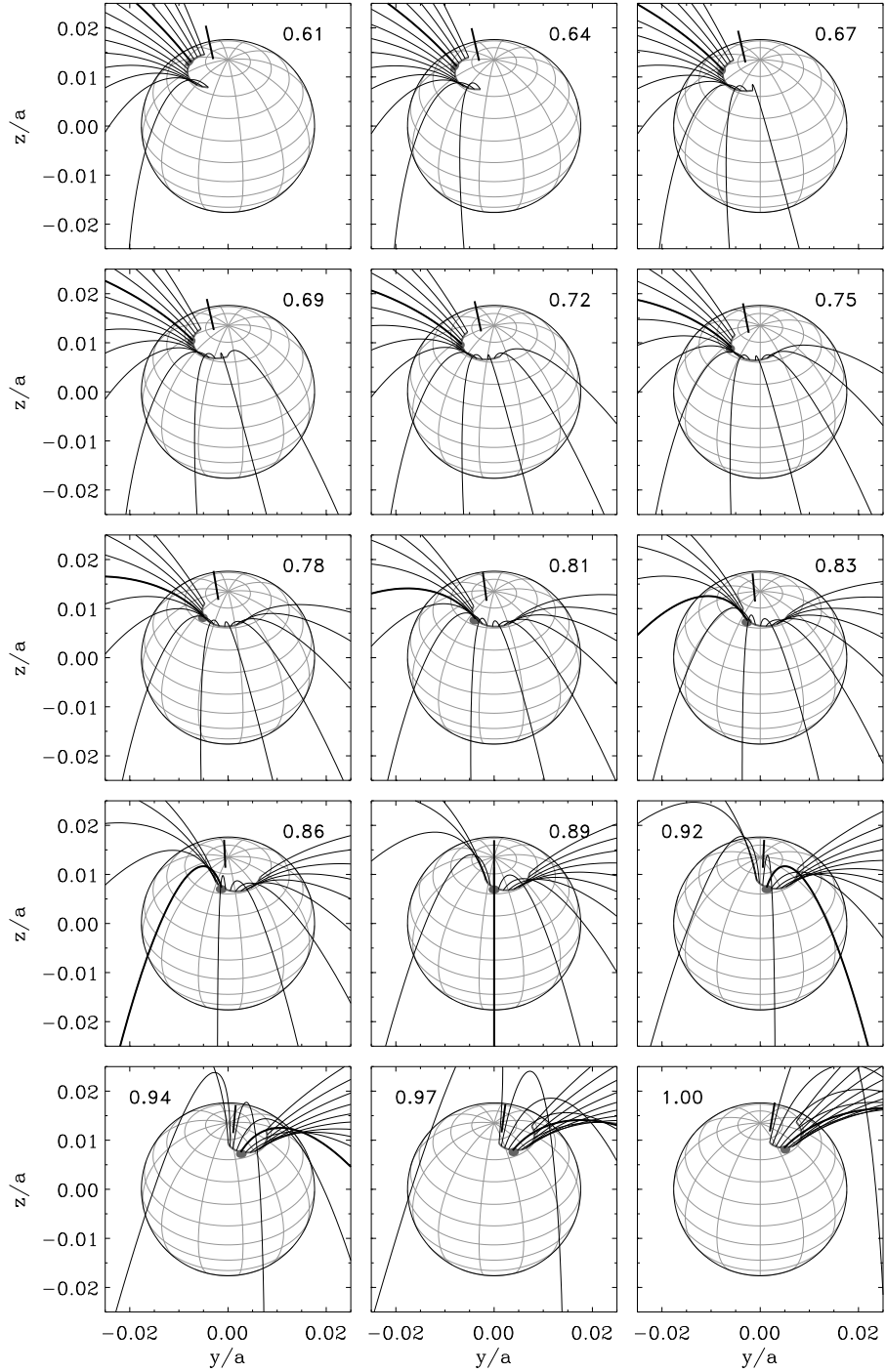}
\caption{Similar to Figure~8, but showing only the white dwarf and
magnetic field lines for a binary inclination $i=50^\circ$ and binary
phases $\phi =220^\circ , 230^\circ , 240^\circ , \dots , 360^\circ $
($\phi=0.61$--1.0). Magnetic pole is indicated by the short thick line
and a spot is suggestively drawn at the footpoint of the dominant
$\varphi = \psi = 40^\circ $ field line.}
\end{figure}

\end{document}